# Hyperpolarization of *cis*-$^{15}$N,$^{15}$N'-azobenzene by parahydrogen at ultralow magnetic fields

Kirill F. Sheberstov*[1,2], Vitaly P. Kozinenko[3,4], Alexey S. Kiryutin[3,4], Hans-Martin Vieth[5], Herbert Zimmermann[6], Konstantin L. Ivanov[3,4], Alexandra V. Yurkovskaya[3,4]

1. *Institut für Physik, Johannes Gutenberg Universität-Mainz, 55128 Mainz, Germany*
2. *Helmholtz-Institut Mainz, GSI Helmholtzzentrum für Schwerionenforschung, 55128 Mainz, Germany*
3. *International Tomography Center SB RAS, Novosibirsk, 630090, Russia*
4. *Novosibirsk State University, Novosibirsk, 630090, Russia*
5. *Freie Universität Berlin, Fachbereich Physik, 14195 Berlin, Germany*
6. *Department of Biomolecular Mechanisms, Max-Planck-Institut für Medizinische Forschung, 69120 Heidelberg, Germany.*

## Abstract

Development of the methods to exploit nuclear hyperpolarization and search for molecules whose nuclear spins can be efficiently hyperpolarized is an active area in nuclear magnetic resonance. Of particular interest are those molecules that have long nuclear relaxation times, making them to be suitable candidates as contrast agents in magnetic resonance imaging. In this work we present a detailed study of SABRE SHEATH (Signal Amplification By Reversible Exchange in Shield Enabled Alignment Transfer to Heteronuclei) experiments of $^{15}$N,$^{15}$N'-azobenzene. In SABRE SHEATH experiments nuclear spins of target are hyperpolarized by transfer of spin polarization from parahydrogen at ultralow fields during a reversible chemical process. The studied system is complicated, and we are concerned only about a subset of the data, presenting details for the molecules that experience fast chemical exchange at the catalytic complex and thus are involved in polarizing the free azobenzene. Azobenzene exists in two isomers *trans*- and *cis*-. We show that all nuclear spins in *cis*-azobenzene can be efficiently hyperpolarized by SABRE at suitable magnetic fields. Enhancement factors (relative to 9.4 T) reach several thousands of times for $^{15}$N spins and a few tens of times for the $^1$H spins. There are two approaches to observe either hyperpolarized magnetization of $^{15}$N/$^1$H spins, or hyperpolarized singlet order of the $^{15}$N spin pair. We compare these approaches and present the field dependencies of SABRE experiments for them. No hyperpolarization of *trans*-$^{15}$N,$^{15}$N'-azobenzene was observed. The results presented here will be useful for further experiments where hyperpolarized *cis*-$^{15}$N,$^{15}$N'-azobenzene is switched by light to *trans*-$^{15}$N,$^{15}$N'-azobenzene for storing the produced hyperpolarization in the long-lived spin state of the $^{15}$N pair of *trans*-$^{15}$N,$^{15}$N'-azobenzene.

## Introduction

Parahydrogen can be used to hyperpolarize nuclear spins allowing enhancement of signals in nuclear magnetic resonance (NMR) and magnetic resonance imaging (MRI) by up to 5 orders of magnitude[1,2]. This method is applicable when a pairwise hydrogenation reaction occurs; then the singlet spin order (population imbalance between the singlet and triplet states of the spin pair) of parahydrogen can be transferred and redistributed across the nuclear spin system of the target molecule. Of particular interest are reversible hydrogenation reactions happening in catalytic complexes: in this case a dihydrogen molecule $H_2$ that is dissolved in a solution can reversibly bind to a metal centre. It is possible to transfer its nuclear spin order to another target molecule that binds to the same complex simultaneously with the hydrogens. The method is called SABRE (Signal Amplification By Reversible Exchange),[3] and it enables polarizing a wide variety of substrates[4], including important biomarkers that can be used as contrast agents for bio-imaging[5]. Importantly, this technique allows to repeat hyperpolarization for many times with the same target molecules as all the chemical processes are cyclic.

The search for new targets for SABRE is an actual problem, as it is still impossible to predict in advance the molecules that are suitable. The catalytic complexes used for SABRE are usually based on iridium (III), thus the target molecules should be able to reversibly bind to iridium. Therefore, the molecule must contain an electron-donating heteroatom, such as nitrogen. There are currently more than a hundred of various compounds known that can be hyperpolarized by SABRE, and most of them contain a nitrogen atom[6].

The most efficient way to transfer the spin order of parahydrogen to the target molecule exploits a coherent transfer through the *J*-coupling network; this process depends strongly on the magnetic field at which SABRE is performed. Heteronuclear spins such as $^{15}N$ and $^{13}C$ can be hyperpolarized by SABRE in zero- to ultralow-fields (ZULF). In ZULF the difference between the Larmor frequency of protons and the heteronuclear spin is comparable to the *J*-coupling between them, making spontaneous coherent polarization transfer through *J*-couplings possible. This method is called SABRE SHEATH[7,8] (Shield Enables Alignment Transfer to Heteronuclei).

Relaxation times of $^{15}N$ and $^{13}C$ spins are typically much longer than those of $^1H$ spins, allowing to store the hyperpolarization up to several minutes. But another possibility is to store hyperpolarization in a so-called long-lived state (LLS)[9], a state that is immune to the most common relaxation mechanisms. The longest relaxation times are observed for singlet order formed within pairs of $^{15}N$ spins[10–12] or $^{13}C$ spins[13], allowing to store nonequilibrium spin states for tens of minutes in liquids at room temperature. The possibility to store hyperpolarization expands the field of applications and is causing intense interest in the development of SABRE experiments producing LLS[14–16]. Here we report on SABRE SHEATH experiments with $^{15}N,^{15}N'$-azobenzene (ABZ), a molecular target with outstanding LLS properties.

ABZ exists in two isomeric forms, namely *trans*-ABZ and *cis*-ABZ, and the prevailing form in solution can be controlled by light (see Figure 1a). Recently, it was discovered that the singlet state of the $^{15}N$ spin pair in *trans*-$^{15}N,^{15}N$-ABZ is long-lived, which enables sustaining of non-thermal spin order up to 50 minutes even at high magnetic field (>10 T)[12]. Illustration of the SABRE process of ABZ is shown in Figure 1b.

The possibility to hyperpolarize *cis*-$^{15}N,^{15}N'$-ABZ with SABRE was reported previously[4], however it was a large scale study of many compounds and information about SABRE of $^{15}N,^{15}N'$-ABZ was limited, even the enhancement level was not determined. Here we present a detailed study on SABRE SHEATH of $^{15}N,^{15}N'$-ABZ and compare experiments, in which hyperpolarized magnetization of $^{15}$N and $^1$H is observed to those experiments, where hyperpolarized singlet order of the $^{15}$N spins pair is observed.

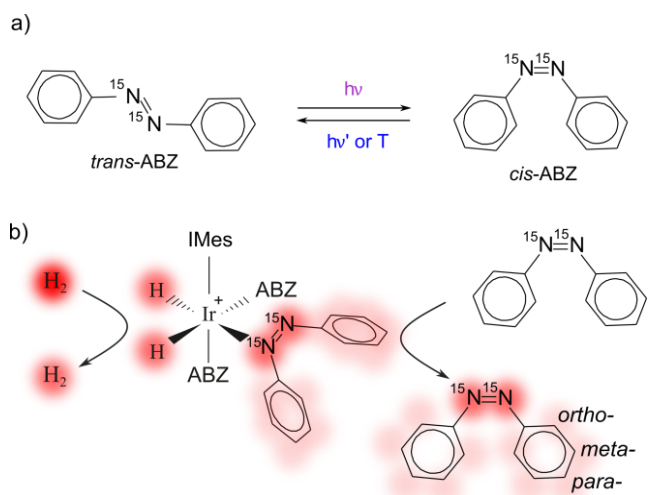

Figure 1. (a) Chemical structure and photochromic property of azobenzene (ABZ). (b) Scheme of SABRE with cis-ABZ showing polarization transfer from a parahydrogen molecule to cis-ABZ within a reversibly formed Ir-Imes activated complex.

**Methods and experimental details**

The spin system of cis-$^{15}$N,$^{15}$N'-ABZ includes 12 coupled spins; all of them can be hyperpolarized by SABRE. There are slightly different experimental protocols allowing different ways to observe the hyperpolarization. These methods are presented in Figure 2. Most of the steps are identical: the sample containing $^{15}$N,$^{15}$N'-ABZ and precatalyst was placed in a magnetic shield; the required ultralow field was set up using a set of coils, and the parahydrogen gas was bubbled through the solution. The bubbling time $\tau$ was chosen according to the relaxation time of the hyperpolarized nuclear spins. In our experiments it was usually set to 10 s. The supply of gas was then stopped by immediate equilibration of pressure at the inlet and outlet of the NMR tube; then the magnetic field was instantaneously turned on up to 25 uT. Such an instantaneous jump of the magnetic field is not necessary, but it allowed better reproducibility of experiments. Then the sample was transferred inside the NMR spectrometer (9.4 T, 400 MHz $^1$H Larmor frequency) where different detection sequences were applied.

The first possibility is to apply a hard 90º pulse (Figure 2a), detecting magnetization of the hyperpolarized spins. This corresponds to the most common SABRE SHEATH experiment. Because the spin system of $^{15}$N,$^{15}$N'-ABZ contains both $^{15}$N and $^1$H spins, detection can be performed either by using the $^{15}$N or the $^1$H channel.

The second possibility is to detect hyperpolarized singlet order of the $^{15}$N spin pair in cis-$^{15}$N,$^{15}$N'-ABZ. It is worth noting that the two $^{15}$N spins in $^{15}$N,$^{15}$N'-ABZ are close to magnetic equivalence, the eigenstates of this pair are close to the singlet and triplet states, hence in most cases singlet order is not observable after applying the usual hard NMR pulses. Special methods were developed to convert singlet order into magnetization; here we use a method that is called spin induced level crossing (SLIC)[17]. SLIC pulses were adjusted to convert singlet order of the $^{15}N$ spin pair of $^{15}N,^{15}N'$-ABZ into observable magnetization[12,18]. SLIC can be used to convert the singlet order into either $^{15}N$ magnetization or $^1H$ magnetization of *ortho*-protons in $^{15}N,^{15}N'$-ABZ. Consequently, the SLIC pulse was applied either at the $^{15}N$ resonance frequency (~525 ppm for *cis*-$^{15}N,^{15}N'$-ABZ) or at the $^1H$-*ortho* frequency (~6.85 ppm for *cis*-$^{15}N,^{15}N'$-ABZ). The nutation frequency matched the $^1J_{NN}$-coupling, which is ~22 Hz, the pulse duration was set to 0.3 s.

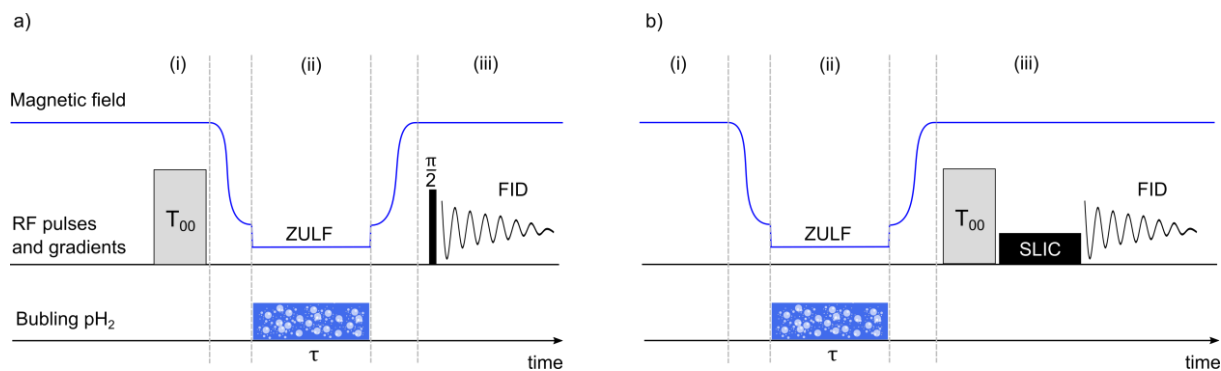

*Figure 2. Experimental protocols. (a) Field cycling in a SABRE SHEATH experiment with consecutive (i) destruction of equilibrium magnetization at the high field of the NMR spectrometer, followed by transfer of the sample into the magnetic shield, (ii) bubbling parahydrogen at ZULF conditions in a variable magnetic field for a time $\tau$, and (iii) backward transfer of the sample and excitation of the NMR signals by a hard pulse. (b) Field cycling in a SABRE SHEATH experiment to observe hyperpolarized singlet order of the $^{15}N$ spin pair in cis-$^{15}N,^{15}N'$-ABZ. In this case, excitation is performed by using a SLIC pulse (see in the text). The $T_{00}$ block indicates a filter based on 90° pulses with pulsed field gradients; for details see reference[19].*

Chemical shifts of $^{15}N$ are referenced on the Ξ-scale utilizing the absolute Larmor frequency of TMS and $\Xi_N$ = 10.1329118 for liquid ammonia to obtain the absolute frequency for 0 ppm of the $^{15}N$ spins[20].

Parahydrogen was enriched up to 90% using a Bruker parahydrogen generator, the pressure inside NMR tube was set to 3 bar. The bubbling time $\tau$ was optimized by measuring the "build up curve", where the intensity of the hyperpolarized signals is plotted as function of $\tau$ (not shown here). After bubbling in a magnetic field of 500 nT for 30 s, the intensity reached a plateau with the level of hyperpolarization about 1.7 times higher as compared to a bubbling time of 10 s. From here on, we fixed the bubbling period to $\tau = 10$ s - this made it possible to achieve a high degree of polarization and, at the same time, moderately consume parahydrogen. Further details of the experimental setup are given in reference[21].

The samples consisted of 50 mM of $^{15}N,^{15}N'$-ABZ and of 3 mM of the precatalyst, [Ir(Imes)(1,5-cycloctadiene)]Cl dissolved in deuterated methanol. The sample was irradiated by near UV light (~380 nm) to convert *trans*-$^{15}N,^{15}N'$-ABZ into *cis*-$^{15}N,^{15}N'$-ABZ. After irradiation, the *cis*-$^{15}N,^{15}N'$-ABZ to *trans*-$^{15}N,^{15}N'$-ABZ ratio reached 1:1. Next, the parahydrogen gas was bubbled through the solution to activate the [Ir(Imes)(1,5-cycloctadiene)]Cl complex. This process was performed inside the NMR spectrometer and monitored by $^1H$ NMR. The activation of the complex took up to 10 minutes without addition of any substrates except for $^{15}N,^{15}N'$-ABZ. The temperature in experiments was set to about 25° C.

The precatalyst, [Ir(Imes)(1,5-cycloctadiene)]Cl, was synthesized using the procedure described in reference[22] and $^{15}N,^{15}N'$-ABZ enriched by $^{15}N$ was synthesized as described in reference[18]. Deuterated methanol-d$_4$ (99.8% atom %D) was purchased from Carl Roth.

**Results and discussion**

Polarization transfer in the spin network of $^{15}N,^{15}N'$-ABZ is a sophisticated process. Hyperpolarization propagates through the strongly coupled nuclei allowing to polarize remote spins. As we show further, the protons in the phenyl rings of $^{15}N,^{15}N'$-ABZ can be polarized only in case if SABRE is performed at ZULF. This is similar to previous studies of metronidazole where remote spins are hyperpolarized through the *J*-coupling network[23]. Also, different spin orders are polarized, of particular interest are $^{15}N$ magnetization and singlet order of the $^{15}N$ spin pair. The ability to hyperpolarize singlet order is reminiscent of the performance of SABRE SHEATH methods in diazarines[15,16].

The large size of the spin system makes it difficult to model the experiments: the full spin system in the complex with 3 $^{15}N,^{15}N'$-ABZ molecules contains 38 coupled spins. Nevertheless, some of the details can be understood from exploring the magnetic field dependencies of SABRE and by modeling the process with simplified spin systems.

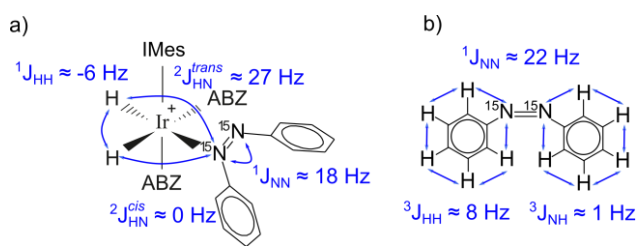

*Figure 3. J-couplings responsible for coherent polarization transfer at ultra-low fields in the Ir-Imes complex (a) and in cis-$^{15}N,^{15}N'$-ABZ (b).*

The relevant *J*-couplings in the Ir-Imes complex with $^{15}N,^{15}N'$-ABZ are shown in Figure 3a, while the *J*-couplings in *cis*-$^{15}N,^{15}N'$-ABZ are shown in Figure 3b. The values for free *cis*-$^{15}N,^{15}N'$-ABZ are taken from reference[18], while the *J*-couplings in Ir-Imes complex were determined from the line splittings in $^1H$ and $^{15}N$ NMR spectra. These spectra were acquired after bubbling parahydrogen in high field at 5 °C, which slowed down the exchange and made the splitting visible (not shown here).

Distinguishment between *cis* and *trans* $^2J_{HN}$-couplings is made in accordance to previous studies of $^2J_{HN}$-couplings in Rh-based complexes[24]. Interestingly, the $^1J_{NN}$-coupling in *cis*-$^{15}N,^{15}N'$-ABZ is different for the free *cis*-$^{15}N,^{15}N'$-ABZ (~22 Hz) and for the molecule bound in equatorial position (~18 Hz). This can be explained by the fact that electron density is shifted from the attached site towards the metal centrum.

**Magnetization formed by SABRE SHEATH**

A typical $^{15}N$ spectrum obtained after SABRE SHEATH of $^{15}N,^{15}N'$-ABZ at 500 nT (protocol of Figure 2a) is shown in Figure 4. In this figure the strongest signal near 525 ppm corresponding to the free *cis*-$^{15}N,^{15}N'$-ABZ is cut. There are several other hyperpolarized signals, which otherwise are invisible in the thermally equilibrium spectrum. All $^{15}$N nuclei in the sample solution belong to $^{15}$N,$^{15}$N'-ABZ, so all these signals are attributed to different signals of either *cis*- or *trans*-$^{15}$N,$^{15}$N'-ABZ attached in equatorial or axial sites of the Ir-Imes complex. Detailed assignment of these signals was not our goal here, but we determined the signals corresponding to *cis*-$^{15}$N,$^{15}$N'-ABZ attached to Ir-Imes in equatorial position. These signals are highlighted in Figure 4. They are broadened due to chemical exchange. The signal at ~420 ppm corresponds to the $^{15}$N position coordinated with the Ir, whereas the remote $^{15}$N nitrogen has almost the same chemical shift as in the free *cis*-$^{15}$N,$^{15}$N'-ABZ. The assignment was confirmed by correlation experiments showing that there is $^2J_{HN}$-coupling between the attached $^{15}$N and one of the parahydrogen protons in the Ir-Imes complex as well as correlation between the $^{15}$N signals.

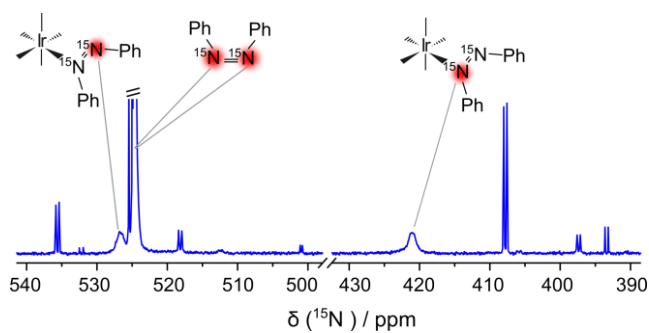

*Figure 4. Fragments of the $^{15}$N NMR spectrum showing signals of $^{15}$N,$^{15}$N'-ABZ in Ir-Imes complexes, polarized by SABRE SHEATH at 500 nT. The phase of the spectrum was adjusted to have absorptive signals. The experimental protocol is shown in Figure 2a.*

Only the free form of *cis*-$^{15}$N,$^{15}$N'-ABZ is hyperpolarized as can be concluded from comparison of the thermal and the hyperpolarized spectra shown in Figure 5. Probably, absence of the hyperpolarization for the free *trans*-$^{15}$N,$^{15}$N'-ABZ can be explained by steric restrictions For the shown spectrum (Figure 4 and Figure 5a) the enhancement factor was ε≈3000. It was calculated as ratio between the integral intensity of the SABRE SHEATH spectrum and the intensity observed in the thermally polarized spectrum. The same sample was used to acquire both spectra, but the thermally polarized spectrum was averaged over 128 acquisitions to get an acceptable signal-to-noise ratio (SNR).

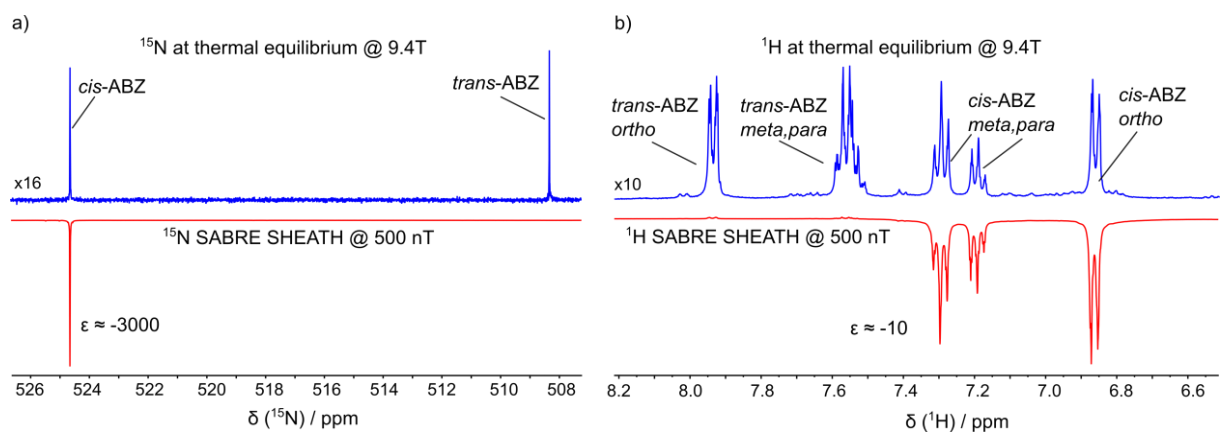

*Figure 5. $^{15}$N NMR spectra (a) and $^1$H NMR spectra (b) showing comparison of $^{15}$N,$^{15}$N'-ABZ signals in thermal equilibrium (blue, absorption peaks) and after SABRE SHEATH hyperpolarization at 500 nT (red, emission peaks). The phases in SABRE SHEATH spectra were set with respect to corresponding thermal spectrum. The $^{15}$N NMR spectrum at thermal equilibrium shown in (a) was averaged for 128 times. Otherwise, all the compared spectra were acquired with identical acquisition and processing parameters. The experimental protocol for the SABRE experiments is shown in Figure 2a.*

$^1$H signals of cis-$^{15}$N,$^{15}$N'-ABZ were hyperpolarized as well (Figure 5b). The enhancement factors reached a value of ε~10 for each of the proton groups. This is about two orders of magnitude less than the enhancement factors for the $^{15}$N signal. The difference can be partially explained by the fact that $^1$H spins have a ~10 times higher absolute value of the gyromagnetic ratio, and therefore thermally polarized $^1$H spins give a ~10 times stronger spectrum. Another order of magnitude may be explained by the weak *J*-couplings between $^1$H and $^{15}$N spins, making polarization transfer to be less efficient. Another possible explanation is that the relaxation times of $^1$H and $^{15}$N spins are substantially different at 500 nT, but they were not measured here.

The integral intensities of the hyperpolarized signals correspond to the stoichiometric ratio of protons in $^{15}$N,$^{15}$N'-ABZ, when SABRE SHEATH is performed at 500 nT. Also, in this case, the detected spin order of protons corresponds to the integral magnetization which changes when SABRE is performed at another magnetic field.

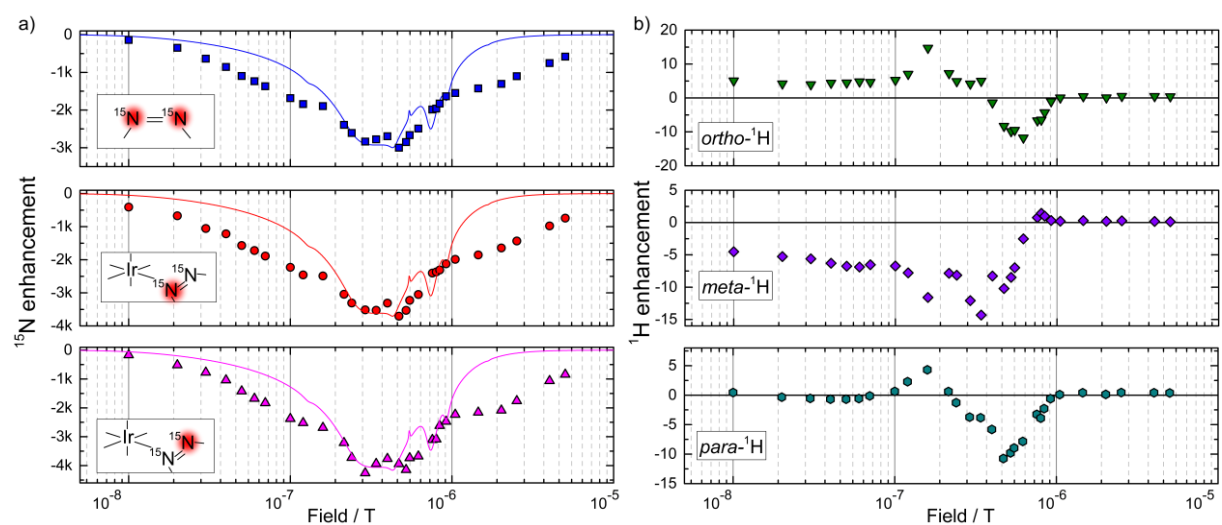

*Figure 6. Field dependence of $^{15}$N NMR signals (a) and $^1$H NMR signals (b) hyperpolarized by SABRE SHEATH. The experimental values of the signal integrals are shown by scatter plots; the calculation results are shown by the solid lines. Details of the calculations are given in the text. All the functions are normalized by the integral intensity of the corresponding signal intensity at thermal equilibrium. Note that the Y-scale is different for each subplot. Top row in (a) shows the data for free cis-$^{15}$N,$^{15}$N'-ABZ. Middle row in (a) shows the data for the $^{15}$N spin of the cis-$^{15}$N,$^{15}$N'-ABZ complex in equatorial position that is directly attached to Ir-Imes. Bottom row in (a) shows the data for the $^{15}$N position of cis-$^{15}$N,$^{15}$N'-ABZ remote from the Ir-Imes complex. Top row in (b) shows data for the ortho-$^1$H spins in cis-$^{15}$N,$^{15}$N'-ABZ. Middle row in (b) shows data for the meta-$^1$H spins in cis-ABZ. Bottom row in (b) shows data for the para-$^1$H spins in cis-$^{15}$N,$^{15}$N'-ABZ. The experimental protocol is shown in Figure 2a.*

The field dependence of the enhancements produced by SABRE SHEATH is shown in Figure 6. Let us first consider the $^{15}$N signals (Figure 6a). The maximal enhancement for the free cis-$^{15}$N,$^{15}$N'-ABZ was achieved in the field range of 200 ÷ 500 nT (Figure 6a, top row). The field dependences of $^{15}$N signals corresponding to cis-$^{15}$N,$^{15}$N'-ABZ in equatorial position of the Ir-Imes complex behave similarly with the signal corresponding to the free cis-$^{15}$N,$^{15}$N'-ABZ (compare rows in Figure 6a). This appearance may be explained by the fast chemical exchange of cis-$^{15}$N,$^{15}$N'-ABZ between its free form and the form bound in equatorial position: by the moment when the sample arrives inside the NMR spectrometer (stage iii in Figure 2a) all the cis-$^{15}$N,$^{15}$N'-ABZ molecules might have experienced the chemical exchange for many times and the observed polarization of $^{15}$N spins in the complex is secondary. It is formed by the hyperpolarized molecules that came from the bulk.

The enhancement factors ε for the $^{15}$N signals in the complex are difficult to measure because it would take too much time to acquire thermally polarized signals for them. Therefore, the presented enhancements were estimated using the following procedure: first, the amount of complexes with cis-$^{15}$N,$^{15}$N'-ABZ was determined from the $^1$H spectrum. To do that we integrated the signals of cis-$^{15}$N,$^{15}$N'-ABZ and the signals corresponding to hydrides at ~ -20 ppm in the $^1$H NMR spectrum of thermally polarized sample. This was done for a sample pressurized with 3 bar (same as in SABRE experiments) of thermally polarized H$_2$ gas. We assume that the $^1$H NMR signal of the hydrates at around -20 ppm can be used to quantify the amount of the Ir-Imes complex with cis-$^{15}$N,$^{15}$N'-ABZ bound in the equatorial position. The ratio between free cis-$^{15}$N,$^{15}$N'-ABZ with respect to cis-$^{15}$N,$^{15}$N'-ABZ bound in equatorial position determined in such a way was found to be about 34. This ratio was then multiplied by the ratio of the integrals in the $^{15}$N spectra of the hyperpolarized signal with respect to the signal of the free cis-$^{15}$N,$^{15}$N'-ABZ in thermal equilibrium and additionally multiplied by a factor of 2 due to stoichiometry. This gave us the enhancements shown in Figure 6a.

This field dependence of $^{15}$N signals is reminiscent of the analogous field dependencies for other substrates, for example, for $^{15}$N-enriched pyridine[21]. A good agreement was achieved between experiments and calculated curves shown as the solid lines in Figure 6a. The experimental function is somewhat broader, probably, due to the presence of additional spins in the system that were not considered in the simulation or due to limited lifetime of active complex where actual polarization transfer happens. The calculation was done for a simplified spin system with just 6 spins: 2 $^1$H spins coming from the parahydrogen and 4 $^{15}$N spins, corresponding to 2 cis-$^{15}$N,$^{15}$N'-ABZ molecules in the equatorial position of the Ir-Imes complex. The calculation was done using the program iRelax available online;[25] the detailed description of the calculation can be found in reference[21]. We note that this calculation does not consider the spin dynamics of the polarization transfer, nor the lifetime of the complex. First, the projection of the initial singlet state of parahydrogen onto the eigen states of the full spin system is calculated. All coherences in the obtained density matrix are set to zero and then the Liouville bracket of the obtained state with the $\hat{I}_z$ operator of all $^{15}$N spins is taken. Interestingly, when larger spin systems with several $^1$H spins of cis-$^{15}$N,$^{15}$N'-ABZ were calculated, they showed poor agreement between experimental and calculated data. We believe that this discrepancy is due to the weak polarization levels observed for the $^1$H spins. Hence, the hyperpolarization of the $^{15}$N magnetization in cis-$^{15}$N,$^{15}$N'-ABZ is not strongly affected by the presence of $^1$H spins.

The magnetic fields that were applied inside the magnetic shield (Figure 2a, stage ii) were always parallel to the field inside the NMR spectrometer. The observed sign of the $^{15}$N signal of the free cis-$^{15}$N,$^{15}$N'-ABZ in SABRE SHEATH experiments was always opposite to the thermally polarized one. This observation confirms that the sign of the $^1J_{HH}$-coupling between the parahydrogen protons in the Ir-Imes complex is negative: simulation with a positive $^1J_{HH}$-coupling inverts the sign of $^{15}$N polarization.

This is also in agreement with the values for similar complexes, but with different substrates[26].

The field dependence of the $^1$H spin hyperpolarization in cis-$^{15}$N,$^{15}$N'-ABZ is shown in Figure 6b. Typically, $^1$H spins of different substrates are hyperpolarized by SABRE in a few mT magnetic field[26]. However, the $^1$H spins of cis-$^{15}$N,$^{15}$N'-ABZ were hyperpolarized by SABRE only in the ZULF regime. This means that they cannot be directly polarized by the parahydrogen protons in the Ir-Imes complex, but instead their hyperpolarization arises as result of polarization transfer from the hyperpolarized $^{15}$N spins. Calculation of the field dependencies of the $^1$H spins should take into account propagation of the polarization averaged over the lifetime of the complex; such simulations go beyond the scope of this work. There are two fields at which all curves of the $^1$H spins have an extremum: at 150 nT and ~400 nT. Data for the *ortho*-protons look similar to the data of *para*-protons. Both groups change sign of their polarization in the field range from 150 to 300 nT. When the field is lowered to almost zero no net magnetization can be formed. However, some hyperpolarization is still observed; it corresponds to antiparallel orientation between *ortho*- and *meta*-protons. Upon increasing the magnetic field beyond 1 uT no hyperpolarization is not formed.

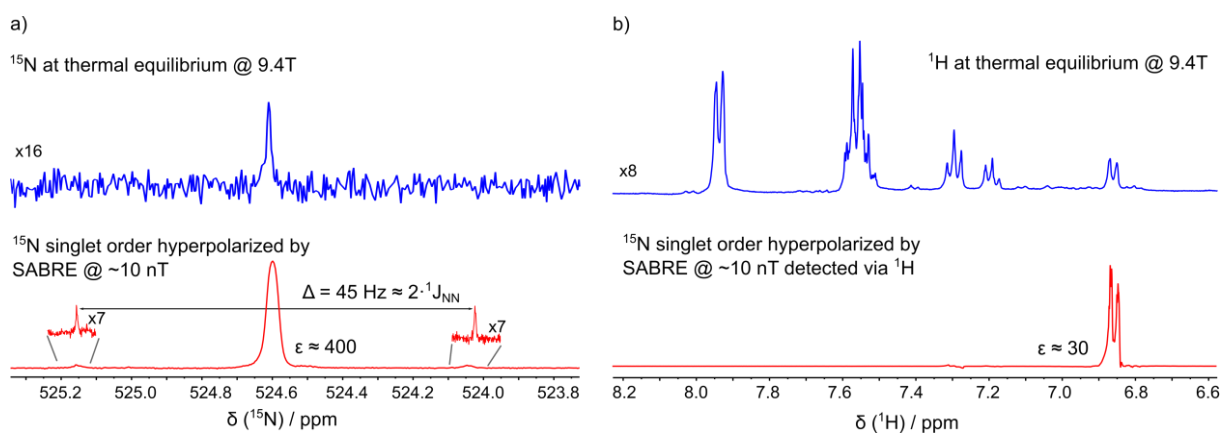

*Figure 7.* $^{15}$N NMR spectra (a) and $^1$H NMR spectra (b) showing comparison of $^{15}$N,$^{15}$N'-ABZ signals in thermal equilibrium (blue) and after SABRE SHEATH hyperpolarization at 10 nT of the singlet order of the $^{15}$N spin pair (red). The thermal spectrum in (a) was acquired in 4 scans with proton decoupling. This narrowed the line and increased SNR. Hyperpolarized $^{15}$N spectrum exhibits outer singlet-triplet coherences. The experimental protocol for the SABRE experiments is shown in Figure 2b.

## $^{15}$N singlet order formed by SABRE SHEATH

Using the experimental scheme shown in Figure 2b, we acquired NMR spectra detecting only the singlet order of the $^{15}$N spin pair of free cis-$^{15}$N,$^{15}$N'-ABZ. Before the SLIC excitation[17], a so-called $T_{00}$ filter[19] was applied to destroy all spin orders except for the singlet order. The following SLIC pulse converted hyperpolarized singlet order of the $^{15}$N spin pair into either $^{15}$N coherence (Figure 7a) or into coherence of the *ortho*-protons of cis-$^{15}$N,$^{15}$N'-ABZ (Figure 7b). Due to the $T_{00}$ filter and the selective excitation, these NMR spectra contain only the hyperpolarized signal. A specific spectral feature was observed in the $^{15}$N spectrum: the main line was surrounded by weak satellites with the line splitting between them equal to approximately two $^2J_{NN}$-couplings. These satellites correspond to enhanced outer singlet-triplet coherences that sometimes can be observed in singlet NMR experiments with nearly equivalent spin pairs[27].

These experiments clearly show that apart from the $^{15}$N and $^{1}$H magnetization, singlet order of the $^{15}$N spin pair is populated during SABRE SHEATH experiments. We note that equilibrium population of the singlet order is usually zero at any magnetic field. The observed lines are much stronger than the normal NMR signals.

The field dependence of how the $^{15}$N singlet order is hyperpolarized, was measured via the $^{15}$N channel (Figure 8a) and via the $^{1}$H channel (Figure 8b). Both curves shown in Figure 8 have a very similar appearance, which is expected, as they correspond to two different ways of observing the same hyperpolarized state. Singlet order is formed most efficiently at zero magnetic field; with increasing field the polarization level gradually drops to zero. There is a wide range of magnetic fields ($10^{-5}$ to $10^{-2}$ T) where almost no changes are happening. This dependence is somewhat similar to the SABRE experiments on singlet order of the $^{15}$N spin pair in diazirines[15,16]. There is a substantial formation of singlet order observable up to $10^{-2}$ T. Theis *et al.* explain this feature by finding out that singlet order enrichment happens due to mixing of spin levels that do not depend on the magnetic field[15].

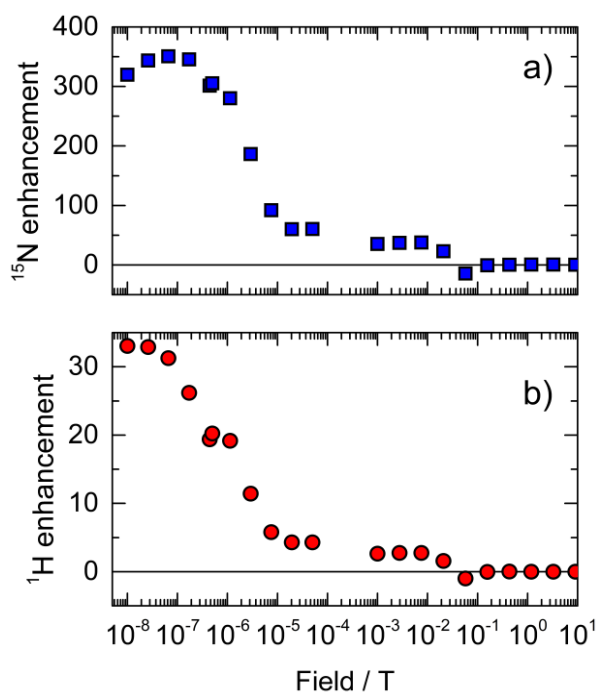

*Figure 8. Field dependence of $^{15}$N NMR signals (a) and $^{1}$H NMR signals (b) corresponding to singlet order of the $^{15}$N spin pair in cis-$^{15}$N,$^{15}$N'-ABZ hyperpolarized by SABRE SHEATH. The experimental protocol for the SABRE experiments is shown in Figure 2b. All the functions were normalized by the integral intensity of the corresponded signal intensity at thermal equilibrium.*

We also measured the field dependence of the relaxation times of the hyperpolarized $^{15}$N singlet order. The lifetime was below 10 s at every field and thus not longer than the corresponding longitudinal relaxation time of $^{15}$N spins in cis-$^{15}$N,$^{15}$N'-ABZ, therefore we do not show this data here. The absence of long-living character is however expected for *cis*-$^{15}$N,$^{15}$N'-ABZ and is in accordance with previous observations[18]. One reason is that *cis*-$^{15}$N,$^{15}$N'-ABZ experience chemical exchange between the free and bound forms, with the later form having in general much shorter relaxation times and active mechanisms that may cause relaxation of the $^{15}$N singlet order. The second reason is that even in the free form the molecular geometry of *cis*-ABZ is not symmetric, making possible for the dipolar interactions of the protons in the phenyl rings with $^{15}$N to drive the relaxation of the $^{15}$N singlet order. This was thoroughly analysed by Stevanato *et al.* for a *cis-* and *trans-* derivative of $^{13}$C$_2$-fumarate[28].

The enhancement factor for the observation of singlet order via the $^{15}$N channel reaches ε~400 and ~30 via the $^1$H channel. A factor of 10 between these values is expected, because $^1$H spins in thermal equilibrium have an approximately 10 times higher polarization than $^{15}$N spins.

**Conclusions**

In this work we studied SABRE SHEATH experiments of $^{15}$N,$^{15}$N'-ABZ and have shown that this molecule is a suitable substrate. There is no need to add other substrates to the solution to activate the [Ir(Imes)(1,5-cycloctadiene)]Cl precatalyst. Addition of *cis*-ABZ and bubbling of hydrogen is enough. Only *cis*-$^{15}$N,$^{15}$N'-ABZ, but not *trans*-$^{15}$N,$^{15}$N'-ABZ, is hyperpolarized in its free form. On the other hand, we observe very complicated $^{15}$N spectra hyperpolarized by SABRE SHEATH. There are many hyperpolarized signals corresponding to $^{15}$N,$^{15}$N'-ABZ bound to Ir-Imes, and further research is necessary to make the proper assignment of these complexes and to measure their properties. Here, we focussed on studying in detail SABRE SHEATH of free *cis*-$^{15}$N,$^{15}$N'-ABZ and of *cis*-$^{15}$N,$^{15}$N'-ABZ bound to the equatorial position of the Ir-Imes complex.

Our study provides details about optimal conditions to perform SABRE SHEATH experiments with *cis*-$^{15}$N,$^{15}$N'-ABZ. Further interest in this molecule is caused by the presence of a long-lived state in *trans*-$^{15}$N,$^{15}$N'-ABZ. The lifetime of the singlet order of the $^{15}$N spin pair in *trans*-$^{15}$N,$^{15}$N'-ABZ reaches tens of minutes even at high magnetic fields (>10 T)[12], thus making it interesting for high-field MRI applications. Experiments combining SABRE hyperpolarization of *cis*-$^{15}$N,$^{15}$N'-ABZ, its photoswitching to *trans*-$^{15}$N,$^{15}$N'-ABZ and storage of the hyperpolarization in the LLS of the $^{15}$N spin pair are currently undergoing in our laboratory.

Here we compared SABRE SHEATH experiments where the $^{15}$N magnetization is observed (Figure 2a) to those where the $^{15}$N singlet order is observed (Figure 2b) in *cis*-$^{15}$N,$^{15}$N'-ABZ. The spectral enhancement is in the latter case reduced by a factor of about 10. The optimal field range to polarize $^{15}$N and $^1$H magnetization lies between 200 and 500 nT, while the optimal magnetic fields to polarize singlet order of the $^{15}$N spin pair are in the range between -100 to 100 nT, though even at relatively high magnetic fields of $10^{-2}$ T some hyperpolarized singlet order can still be found.

The achievable hyperpolarization level at optimal conditions which is observed via the $^1$H channel is highest for the experiments, where singlet order of the $^{15}$N spin pair is observed. The enhancement ε ~ 30 means that for achieving the same SNR in the thermally polarized $^1$H NMR spectrum of *cis*-$^{15}$N,$^{15}$N'-ABZ it would take an about 900 times longer acquisition time. Detection of the hyperpolarized singlet order of the $^{15}$N spin pairs via $^1$H channel seems to be an interesting perspective for several reasons. First, the detection via the $^1$H channel is more sensitive by ~ 2 orders than the detection via the $^{15}$N channel, even the difference in enhancement factor not always compensate for this. Second, this channel is preferable for potential MRI applications, as most of MRI instruments are adjusted to observe the $^1$H magnetization. Finally, we note that the scheme shown in Figure 2b achieves highly selective excitation of the hyperpolarized signals of free *cis*-$^{15}$N,$^{15}$N'-ABZ allowing observation of only the hyperpolarized component, thus making *cis*-$^{15}$N,$^{15}$N'-ABZ an interesting molecule, that can be used in singlet-contrast MRI experiments[29].

## Acknowledgments

We acknowledge the Russian Science Foundation for the grant #20-63-46034. KFS has received funding from the European Union's Horizon 2020 research and innovation programme under the Marie Skłodowska-Curie grant agreement No 766402.

## Keywords

Parahydrogen, NMR, SABRE, azobenzene, ZULF